\input amstex.tex
\documentstyle{amsppt}
\magnification\magstep1
\topmatter
\title Trace extensions, determinant bundles, and gauge group cocycles
\endtitle
\author Joakim Arnlind and Jouko Mickelsson \endauthor
\affil Mathematical Physics, Royal Institute of Technology \\
SE-106 91, Stockholm, Sweden. e-mail jouko\@ theophys.kth.se \endaffil
\date May 14, 2002; revised September 3, 2002 \enddate
\endtopmatter
\advance\vsize -2cm
\magnification=1200
\hfuzz=30pt

\NoBlackBoxes
\document

ABSTRACT  We study the geometry of determinant line bundles associated
to Dirac operators on compact odd dimensional manifolds. Physically,
these arise as (local) vacuum line bundles in quantum gauge theory.
We give a simplified derivation of the commutator anomaly formula
using a construction based on noncyclic trace extensions and
associated nonmultiplicative renormalized determinants.

\redefine\det{\text{det}}
\define\TR{\text{TR} }
\define\tr{\text{tr}}

\define\res{\text{res}}

\vskip 0.3in

1. INTRODUCTION

\vskip 0.3in

The use of noncyclic renormalized traces has been implicit in
perturbative quantum field theory for a very long time. The 1-loop
Feynman diagrams can be thought of as traces of pseudodifferential
operators (PSDO's) which are composed of  products of inverses of (free) Dirac
operators or Laplacians and multiplication operators defined by external
fields. However, typically Feynman diagrams
are ultraviolet divergent which reflects the fact that the
pseudodifferential operator is not of trace class. Instead, one has to
introduce an infinite renormalization in the form of cutoff dependent
subtractions, dimensional regularization, or some other more or less
standard method. This means that one is extending  the trace
to a family of nontraceclass operators.

For renormalized traces the basic property $\tr(AB)=\tr(BA)$ is
usually lost. Likewise, if we try to define a determinant by the use of
the formula $\det(e^A) = e^{\tr(A)},$ the renormalized determinant
is plagued by the multiplicative anomaly, $\det(AB) \neq \det(A) \cdot
\det(B).$ The multiplicative anomaly has been extensively studied in
mathematics  literature, [KV],[F], [D], [O]; see also  [PR]
for related matter on renormalized traces and the geometry of determinant line bundles
associated to families of Dirac operators. 
There has been a discussion in physics literature on the role of the
multiplicative anomaly for computations of effective actions;
see e.g. [CZ], [ECZ], [EFVZ].  
 
Here we shall discuss  another application of noncyclic trace
extensions in quantum field theory. We shall study the determinant
bundles and their relation to  hamiltonian gauge anomalies
along the lines of the paper [MR]. The important difference as
compared to [MR] is that we shall extensively take advantage of the
special features of the trace calculus for pseudodifferential
operators. The setting differs from [PR]; here we are dealing with
line bundles associated to zero order pseudodifferential operators
(signs of Dirac hamiltonians), in contrast to positive order
operators  and the determinant calculus is geared to deal with
operators having principal symbol equal to a unit matrix; this
leads to some changes and simplifications in the calculations.

In [MR] a general abstract
formula for the gauge commutator anomaly was derived, and it was
only later shown that it was equivalent with a \it local anomaly
formula, \rm [LM]. The main result of the present paper is a
construction of the determinant line bundle and the gauge group
action in such a way that it leads directly to a local expression,
including the local Schwinger terms for current commutators,
compatible with standard perturbative QFT calculations.

The results below give a concrete geometric realization of a  
more general property of BRST cocycles in quantum field theory models 
based on noncommutative geometry. As shown in [LMR], the BRST cocycles 
(including commutator anomalies) can typically be expressed as 
generalized traces of sums of
commutators of operators; in the case of PSDO's this leads always 
to local expressions involving integrals of de Rham forms.   

We want to thank the referee for a careful reading of the manuscript
and suggestions leading to improvements in  the original version. 
\vskip 0.3in

2. RENORMALIZED DETERMINANTS

\vskip 0.3in Let $T=1+X$ be a pseudodifferential operator (PSDO) 
acting
on sections of a hermitean vector bundle $E$ over a compact closed
manifold $M$ such that the order of $X$ is less or equal to $-1.$
Recall first the definition of generalized Fredholm determinants
[S], [MR].  If dim$M=d$ then the operator $X$ belongs to the
Schatten ideal $L_p$ of operators in the Hilbert space $L^2(E)$ of
square integrable sections for any $p>d.$ The ideal $L_p$ consists
of all bounded operators $A$ such that $|A|^p$ is a traceclass
operator. Fix an integer $p>d.$ We can define a generalized
determinant by
$$ \det_p(1+X) = \det((1+X)e^{-X + \frac12 X^2 \dots+(-1)^{p-1} \frac{1}{p-1}
X^{p-1}}),\tag2.1$$
where the determinant on the right-hand-side is an ordinary Fredholm
determinant, which converges since the log of the operator argument
is a trace-class operator. This generalized determinant is nonzero if
and only if $1+X$ is invertible, but it is not multiplicative;
instead, one has
$$\det_p((1+X)(1+Y))= \det_p(1+X) \cdot \det_p(1+Y) \cdot
e^{\gamma_p(X,Y)},\tag2.2$$ where $\gamma_p$ is a polynomial of
order $p.$ In the case $p=1$ we have $\det_1=\det,$ which is
multiplicative, for $p=2$ one has $\gamma_2(X,Y)= -\tr(XY).$

The problem with the above determinant is that in general $\det_p(1+X)
\neq \det(1+X)$ even when $X$ is a trace-class operator.

We introduce an \it improved renormalized determinant \rm $\det_{ren}$ such
that
$$\det_{ren}(e^X) = e^{\TR(X)},$$
where $\TR$ is a generalized (noncyclic) trace. The definition of
$\TR$ involves a choice of a positive PSDO $Q$ of positive order
$q.$ We set
$$\TR(A) = \TR_Q(A) = \lim_{z\to 0 }\tr\left(Q^{-z}A -\frac{1}{qz} \res(A)\right),\tag2.3$$
where $\res(A)$ is the Wodzicki residue of the PSDO $A.$ The limit
at $z=0$ gives the finite part of  the singular trace (with poles
at integer points $z.$) The trace $\TR$ has the important property
that it agrees with the ordinary trace $\tr$ whenever $A$ is a
traceclass pseudodifferential operator. This trace has been
studied in detail in [MN], [CDMP]. In particular, they proved the
important formula
$$ \TR_Q [A,B] =  -\frac1q \res([\log|Q|, A]B).\tag2.4$$
Here $\res$ is the Wodzicki residue; it is defined as
$$ \res(A) = \frac{1}{(2\pi)^d} \int_M dx \int_{|p|=1} \tr\,
a_{-d}(x,p),$$ where $a_{-d}$ is the homogeneous term of order
$-d$ in the momenta, in the asymptotic expansion of the (matrix
valued)  symbol of $A.$  

 A particular case of this was used earlier in a
study of QFT anomalies, [M], [LM2]. In that case it was assumed
that the symbols are globally defined, with supports of the $x$
coordinate in a compact subset of $\Bbb R^d$ so that one can
define a cut-off trace
$$\tr_{\Lambda} (A) = \frac{1}{(2\pi)^d} \int_{|p|\leq \Lambda}
\tr\, a(x,p) dx dp\tag2.5$$
which has an asymptotic expansion
$$\tr_{\Lambda} (A) = \sum_{i=1}^k \Lambda^i \alpha_i(A) + \alpha_0(A)
+ \log(\Lambda) \res(A) + \sum_{i=-1}^{-\infty} \Lambda^i \alpha_i(A).
\tag2.6$$
The renormalized trace is then the finite term $\alpha_0(A).$ It
agrees with the earlier definition when we take $Q=-\Delta,$ the
Laplace operator in $\Bbb R^d.$

An important consequence of (2.4) is that the renormalized trace
of a commutator is \it local \rm in the sense that it is given in
terms of integrals of finite number of symbols in the asymptotic
expansion of the pseudodifferential operators involved.

We now set
$$\det_{ren} (1+X) = \det_p(1+X) \cdot e^{\TR(X-\frac12 X^2 +\dots
+(-1)^p\frac{1}{p-1}X^{p-1})}\tag2.7$$
with $p$ any integer strictly larger than $d;$ the value of the
determinant does not depend on the choice of $p.$ From the definition
follows immediately that
$$\det_{ren} (e^X) = e^{\TR(X)}\tag2.8$$
for any pseudodifferential operator $X$ of order $-1.$ In
particular, when ord$X < -d$ we have $\det_{ren} (e^X) =
\det(e^X).$ The reason we prefered to use (2.7) as a definition
instead of the simpler formula (2.8) is that the right-hand-side of
(2.7) is manifestly independent of  the choice of logarithm of $A=1+X$ 
whereas when using (2.8) one has give the (not very difficult) proof
of it.    

\proclaim{Proposition 1} (Multiplicative anomaly). Let $X,Y$ be a pair
of PSDO's of order $-1.$ Then
$$\det_{ren}((1+X)(1+Y)) = \det_{ren}(1+X) \cdot \det_{ren}(1+Y) \cdot
e^{\gamma(X,Y)},$$ where $\gamma(X,Y) = \TR \left(\log((1+X)(1 + Y)) -
\log(1+X)-\log(1+Y)\right)=$ \newline
$\TR\left(\frac12[X,Y]+\frac16[Y,X^2]+\frac16[Y,YX]-
\frac16[X,Y^2]- \frac16[X,YX] + \dots\right).$ 

\endproclaim

\demo{Proof} By direct computation from $\log \det_{ren} (1+X) = \TR\,
\log(1+X).$ Note that we do not need to worry about the convergence of
the logarithmic expansions since only a finite number of the
commutators contribute to the trace. \enddemo

\bf Remark \rm  The $L_p$ determinants  $\det_p$ are continuous in the 
$L_p$ Schatten ideal topology. The renormalized determinant is not
continuous with respect to this topology. Instead, from
Prop. II.3.4. in [D] follows that it is continuous in a Frechet
topology on the pseudodifferential symbols. In the same way, the
correction $\gamma(X,Y)$ is continuous in the Frechet topology; this 
follows in fact directly from the property that it is a residue of a
polynomial in the operators $X,Y;$ all terms of order higher than
dim$\,M$ in the variables $X,Y$ drop out by (2.4) and by the
definition  of the residue.

Since $\gamma(X,Y)$ is a renormalized trace of a sum of
commutators it follows that the logarithm of the multiplicative
anomaly is local.  Note that in the case $d=1,2$ the anomaly
$\gamma$ vanishes identically whereas in the case $d=3$ we get the
simple formula
$$\det_{ren}( (1+X)(1+Y)) = \det_{ren}(1+X)\cdot \det_{ren} (1+Y) \cdot
e^{\frac12 \TR[X,Y] }.\tag2.9$$
This follows from the fact that in three dimensions $\TR[A,B]=0$ if
the sum of orders of $A,B$ is less or equal to $-3,$ by formula (2.4),
whereas in the cases $d=1,2$ the trace $\TR[A,B]$ vanishes for all
operators $A,B$ of order less or equal to $-1.$

\vskip 0.3in

3. THE DETERMINANT BUNDLE

\vskip 0.3in

Let $D$ be the Dirac operator defined in the spinor bundle over an
\it odd dimensional \rm compact closed manifold $M.$ The Hilbert space $H$ of
square integrable sections of the spin bundle has a spectral
decomposition $H= H_+\oplus H_-$ corresponding to $D\geq 0$ and $D<0$ in
the subspaces $H_{\pm}.$ We also allow for the possibility that the
spin bundle is tensored with a trivial complex vector bundle of finite
rank $N.$ The multiplication operators in $H$ are then represented by
smooth $N\times N$ matrix valued functions on $M;$ they are examples
of bounded PSDO's of order zero and their interest to us lies in the
fact that they represent infinitesimal gauge transformation in
Yang-Mills theory. 

Denote by $\epsilon$ the sign of the Dirac operator, i.e., the grading
operator corresponding to the splitting $H=H_+\oplus H_-.$ It is a
PSDO of order zero with the property $\epsilon^2=1.$ The infinitesimal
gauge transformations are contained in an associative algebra $\Cal A$
which consists of bounded PSDO's $a$ such that
$[D,a]$ is bounded and
$[\epsilon,a]$ is order $-1$ for any $a\in\Cal A$ (this is of course a
part of the definition of a spectral triple in noncommutative
geometry, [C]).

Let us define the infinite dimensional Grassmann manifold
$Gr=Gr(H_+\oplus H_-)$ as the set of all subspaces $W\subset H$ with
the property that the orthogonal projection $pr_-:W \to H_-$ is a
PSDO of order $\leq -1.$ 

Let $\Cal G$ be the gauge group consisting of all
smooth functions $M\to U(N);$ 
note that the $\Cal G$ orbit of $H_+$ is
contained in $Gr.$  
The group $\Cal G$ is contained
in the group $GL_{d+}$ of all invertible elements in $\Cal A;$
the notation is introduced in order to remind that elements $g$ of
$GL_{d+}$ have the property that $[\epsilon,g]^d$ is a PSDO of order
$-d$ and is thus contained in the Dixmier ideal $L_{1+}.$

The Grassmannian $Gr$ is a subset of any Schatten Grassmannian $Gr_p$
for $p> d,$ where $Gr_p$ is defined by the condition that $pr_-$
belongs to the Schatten ideal $L_p.$ For this reason the determinant
bundles $DET_p$ studied in [MR] can be restricted to define complex
line bundles over $Gr.$ However, we want to take advantage of the
special properties of $Gr$ is order to get \it local formulas \rm
to describe the geometry and the gauge group action in the determinant
bundle.

The Grassmannian $Gr$ splits to connected components $Gr^{(k)}$
labelled by the Fredholm index $k$ of the projection $pr_+.$
The operator $pr_+$ is Fredholm since $pr_-$ (the orthogonal
projection on $H_-$) is compact. Likewise,
the group $GL_{d+}$ is an union of connected components labelled
by the index of $pr_+ g pr_+: H_+ \to H_+.$

Let $H_+^{(k)}= H_+\oplus V_k$ for $k\geq 0$ and $H_+^{(k)} = H_+\ominus V_k$
for $k<0,$  where $V_k$ is a $k$ dimensional
subspace of $H_-$ (resp. of $H_+$). We denote by $pr^{(k)}_+$ the
projection onto $H_+^{(k)}.$

Following [PS], we define the Stiefel manifold $St$ as the set
of linear isometries $w: H_+  \to H$ such that $pr_+^{(k)}\circ
w -1$  is a  PSDO of order $\leq -2.$
Note that the image $w(H_+)$ is then an element of
$Gr^{(k)}.$ Thus there is a natural projection $\pi:St\to Gr$ defined by
$w\mapsto w(H_+).$

If $w\in \pi^{-1}(W)\in St$ then $w\circ q$, where $q:H_+\to
H_+$ is an invertible pseudodifferential operator, is in the same fiber if and only
if $q-1$ is a PSDO of order $\leq -2.$  Thus $St$ is a principal
bundle over $Gr$ with fiber the group $U^{(2)}(H_+)$ of unitary PSDO's
$q$ with the above property.

The determinant bundle $DET$ over $Gr$ is defined as the complex line
bundle $St\times \Bbb C/\sim,$ where the equivalence relation is
$$(w\circ q, \lambda)\sim (w, \omega(w,q)\lambda),\tag3.1$$
with
$$\omega(w,q)= \det_{ren}(q) \cdot e^{\gamma(pr_+^{(k)}\circ
w,q)},\tag3.2$$
where $w\in St^{(k)}=\pi^{-1}(Gr^{(k)})$ and $\lambda\in \Bbb C.$
This is really an equivalence relation, since
$$\omega(w,qq') = \omega(wq,q')\cdot \omega(w,q) \tag3.3$$
which in turn follows directly from the multiplicative anomaly
relation, Prop. 1.

By the remark at the end of the Section 2,  $\omega(w,q)\equiv
\det_{ren}(q)$  when the dimension $d=1,2,3.$ The determinant
$\det_{ren}$ is actually multiplicative in these dimensions
when the argument $q$ satisfies ord$(q-1) \leq -2.$

The dual determinant bundle $DET^*$ is defined analogously by the
relation
$$(wq,\lambda ) \sim^* (w, \lambda \omega(w,q)^{-1}).$$
A section of the dual determinant bundle is then a complex function
$\psi:St \to \Bbb C$ such that
$$\psi(wq) = \psi(w) \cdot \omega(w,q).\tag3.4$$
A particular section is given by the determinant function itself,
$\psi(w) = \det_{ren}(pr_+^{(k)}\circ w)$ for $w\in St^{(k)}.$

The group $GL_{d+},$ which contains the gauge group $\Cal G,$ acts
naturally on $Gr.$ However, this action does not lift to $St$ and this
leads to group extensions of the type which were discussed in [MR] in
the context of Schatten ideal Grassmannians. Taking advantage
of special features of the PSDO algebra here  we get
simplified local expressions for the cocycles describing the gauge
group extensions.

We first define the set $\Cal E$ consisting of triples $(g,q,\mu)$
where $g\in GL_{d+}, q\in GL(H_+)$ such that  $ pr_+\, g\, pr_+ -q$ is
a PSDO of order $ -2$ and $\mu$ is a smooth $\Bbb C^{\times}$
valued function on $Gr.$ 

The elements of $\Cal E$ act on $DET$ (or on $DET^*$). The action is given
by
$$(g,q,\mu)\cdot(w,\lambda)= (gwq^{-1},\mu(\pi(w))\lambda\alpha(g,q,w)).$$
Here the function $\alpha$ must be chosen such that
$$\frac{\alpha(g,q,wt)}{\alpha(g,q,w)}=
\frac{\omega(w,t)}{\omega(gwq^{-1},qtq^{-1})}\tag3.5$$ in order
that $(g,q)$ maps equivalence classes to equivalence classes. We
denote by $F$ the grading operator corresponding to the
decompositon $H=W\oplus W^{\perp},$ that is, $F$ is the unit
operator in $W=\pi(w)$ and $(-1)$ times the unit operator in the
complement $W^{\perp}.$ A particular solution of (3.5) is given by
$$ \alpha(g,q,w)=   \frac{\det_{ren}(w_+)}{\det_{ren}(gwq^{-1})_+}
\frac{\det_{ren}(\frac12 q^{-1}(a(F_{11} +1)
+bF_{21}))}{\det_{ren}(\frac12(F_{11} +1))} \tag3.6$$ where $w_+ =
pr_+ \circ w$ and we use the block decompositions
$$F= \left( \matrix F_{11} & F_{12}\\ F_{21} & F_{22}\endmatrix\right)
\text{ and } g=\left(\matrix a&b\\c&d\endmatrix\right)$$
corresponding to the fixed polarization  $H=H_+ \oplus H_-.$  In
the case $d=3$ there is a simpler solution:

\proclaim{Lemma 2} In the case of dimension $d=3$ the function
$$\alpha_{d=3} (g,q,w)= e^{-\TR[q, w_+ q^{-1}]}, \tag3.7$$
where $w_+= pr_+ \circ w,$ satisfies (3.5). \endproclaim

\demo{Proof}
Since $w_+ -1$ and $t-1$ are of order $-2$ we have $\omega(w,t)=
\det_{ren}(t).$ On the other hand, the right-hand-side of (3.5) is
given by the multiplicative anomaly,

$$\frac{\omega(w,t)}{\omega(gwq^{-1}, qtq^{-1})} =
\frac{\det_{ren}(t)}{\det_{ren}(qtq^{-1})}.$$ 

Writing $t=e^z$ we get
$$\frac{\det_{ren}(t)}{\det_{ren}(qtq^{-1})}= e^{\TR(qzq^{-1} -z)}
= e^{\TR[q,zq^{-1}]}= e^{\TR[q,(t-1)q^{-1}]}.$$
But the logarithm of the left-hand-side of (3.5) is
$$\TR[q,w_+ q^{-1}] -\TR[q,w_+ tq^{-1}] = -\TR[q,tq^{-1}]
=-\TR[q,(t-1)q^{-1}]$$ since $(w_+ -1)(t-1)q^{-1}$ is of order $-4$ and
therefore its commutator with the bounded operator $q$ has zero
trace. This  proves the statement. One might wonder why we did not use
the multiplicative anomaly formula directly to $\det_{ren}(qtq^{-1}).$ 
The reason is that $q$ is not of type $1 +$ negative order operator
for which the formula in Proposition 1 is valid. 

\enddemo

The action on $DET$ defines a group structure in $\Cal E.$ The
multiplication is given by
$$(g,q,\mu)(g',q',\mu')= (gg',qq',\mu \mu_g' \theta(g,g',q,q';
\cdot))\tag3.8$$
where $\theta$ is a function on $Gr$ defined by
$$\theta(g,g',q,q'; W) = \alpha(g,q,g'w{q'}^{-1})\alpha(g',q',w)
\alpha(gg',qq',w)^{-1}\tag3.9$$
where $W=\pi(w).$ Here $\mu_g$ denotes the translated function
$\mu_g(W)= \mu(g^{-1}W).$ The right-hand-side of (3.9) is invariant
with respect to the transformation $w\mapsto wt$ with $t\in
GL^{(2)},$ as follows from (3.5), and therefore it is indeed a
function of $W=\pi(w).$

  \proclaim{Proposition 3} The subset $N$ of elements
$(1,q,\mu),$ with $q\in GL^{(2)}$ and
$\mu(W)=(\omega(w,q^{-1})\alpha(1,q,w))^{-1}$ (with $W=\pi(w)$) forms the
maximal normal subgroup of $\Cal E$ which acts trivially on $DET$
and therefore the group $\widehat{GL_{d+}}= \Cal E/N$ acts on
$DET.$
\endproclaim
\demo{Proof}
In order that $(g,q,\mu)$ acts trivially on $DET$ it has to act trivially
on the base $Gr$ and therefore $g=1.$ Next
$$\gather (1,q,\mu)\cdot(w,\lambda)=
(wq^{-1},\lambda\mu(\pi(w))\alpha(1,q,w)) \\
\sim(w,\lambda \mu(\pi(w))
\alpha(1,q,w)\omega(w,q^{-1}) )= (w,\lambda)\endgather$$
for all $w\in St$ if and only if
$\mu(\pi(w))\alpha(1,q,w)\omega(w,q^{-1})=1$ for all $w.$ 
Note that this expression depends on $W=\pi(w)$ and not on $w,$ 
as follows from the relation (3.5); in fact, using the solution (3.6) 
we have $\mu\equiv 1.$

\enddemo

The group $\widehat{GL_{d+}}$ is an extension of $GL_{d+}$ by the
normal abelian subgroup consisting of triples $(1,1,\mu),$ i.e.,
by the group $Map(Gr,\Bbb C^{\times}).$ Locally, near the unit
element, the extension can be given by a 2-cocycle $\xi$ on
$GL_{d+}$ with values in $Map(Gr,\Bbb C^{\times}).$ Using the
local section $g\mapsto (g,a,1)$ the 2-cocycle is given by
$$\xi(g,g')(W) = \theta(g,g',q,q';W)\omega(w_+, (aa')^{-1}a'')^{-1}.
\tag3.10$$ 

In the important case $d=3$ we can use the simple formula (3.7) and
obtain
$$\gather \xi_{d=3}(g,g')(W)= \det_{ren} ((aa')^{-1} {a'}') \cdot
\exp\left\{-\frac12 \TR\left([w_+, (aa')^{-1}{a'}'] + \right.\right.
\\
\left.\left. [{a'}'w_+(aa')^{-1} {a'}', {{a'}'}^{-1} ] -[a(g'w{a'}^{-1})_+,
a^{-1}] -[a'w_+,{a'}^{-1}]\right)\right\}.\tag3.11\endgather$$

Using the standard formula
$$[X,Y]= \frac{d^2}{dt ds}|_{t=s=0} e^{tX} e^{sY} e^{-tX} e^{-sY}\tag3.12$$
relating the group multiplication to Lie product, one obtains a
formula for the commutator in the Lie algebra extension
$\widehat{\bold{gl}_{3+}}= \bold{gl}_{3+} \oplus Map(Gr,\Bbb C),$
where $\bold{gl}_{3+}$ is the Lie algebra of $GL_{3+},$

$$[(X,\mu),(Y,\nu)] = ([X,Y], X\cdot \nu - Y\cdot \mu +c(X,Y)).
\tag3.13$$
Here $X\cdot \nu$ denotes the Lie derivative of the function $\nu$ 
defined by the left group action, $X\cdot \nu= \frac{d}{dt}
\nu_{g(t)}|_{t=0}$ for $g(t)=\exp(tX).$ 

In the case $d=3$ we obtain from (3.11)  (setting $g=e^{tX}$ and $g'=e^{sY}$ and
taking the second derivative with respect to $t,s$) the formula
$$c_{d=3}(X,Y)= \TR\left( b_Y c_X - b_X c_Y -[a_X, b_Y w_-] + [a_Y,
b_X w_-]\right) 
\tag3.14$$
for the Lie algebra $2$ cocycle. 
We have  used the same block decomposition for the operators $X,Y$
as we have used for the group elements $g\in GL_{3+}.$  The
cocycle should be a function of $F\in Gr,$ not of the variable $w$
in the Stiefel manifold. This is indeed the case; it follows from
the fact that $w_- -\frac12 F_{21}$ is of order $-3$ and $b_X$ is
of order $-1.$ Thus the difference $b_X w_- - \frac12 b_X F_{21}$
is of order $-4$ and therefore its commutator with any bounded
PSDO has a vanishing trace. We can now write
$$c_{d=3}(X,Y;F) = \TR\left( b_Y c_X - b_X c_Y -\frac12 [a_X, b_Y F_{21}]
 +\frac12 [a_Y, b_X F_{21}] \right).\tag3.15$$

We finish the discussion by giving an important application of formula (3.15) in
gauge theory.

In Yang-Mills theory the variable $F \in Gr$ comes as the sign of
the Dirac hamiltonian coupled to a Yang-Mills potential. It
specifies tha vacuum in the fermionic Fock space parametrized by
an external vector potential $A,$ [MR]. The operators $X,Y$ are
infinitesimal gauge transformations acting on the fermion field.
Writing $F= (D_0 + A)/|D_0+A|,$ where $D_0$ is the free massless
Dirac hamiltonian, acting on right-handed spinors, in three
dimensions and $A= A_k \sigma^k$ is the potential (each component
$A_k$ with $k=1,2,3$ is a smooth function on $M$ with values in a
Lie algebra of a compact Lie group). The $\sigma$'s are the hermitean 
complex $2\times 2$ Pauli matrices. The first two terms on the
right in (3.15) can be written as
$-\frac18\TR(1+\epsilon)[[\epsilon, X], [\epsilon,Y]].$ This
vanishes by the fact that the Pauli matrices (involved in
$\epsilon$) are traceless and by a parity argument, odd powers in
momenta lead to vanishing integrals in the momentum space.

One can expand the symbol of $F=F_A$ in powers of $A,$ denoting
$\bold{p}=p_k \sigma^k,$ as
$$symb(F_A -\epsilon) = \frac{1}{|p|}\frac12  \left( A - \frac{1}{p^2}
\bold{p} A \bold{p}\right) + O(1/|p|^2).$$ We need to take into
account only a the first two terms in the expansion since higher
order terms lead to operators which give vanishing trace in the
commutators in (3.15), see [LM2] for similar calculations. The
final result is:

\proclaim{Theorem 4} The group of smooth gauge transformations
acts through the extension (3.11) on sections of the determinant
bundle $DET$ over $Gr.$ The corresponding Lie algebra action
leads to an extension of the Lie algebra of infinitesimal gauge
transformations by the abelian ideal of smooth functions on $Gr.$
When the point $F\in Gr$ is parametrized by vector potentials
using the sign of the chiral Dirac hamiltonian  and  $d=3$ the Lie
algebra extension is given by the Mickelsson-Faddeev-Shatasvili
cocycle
$$c_{d=3} (X,Y;A) = \frac{i}{24\pi^2} \int_M \tr A [dX, dY].
\tag3.16 $$
\endproclaim
This agrees with [LM2], [M]  and the cohomological arguments in [M2], [F-Sh].

\vskip 0.5in
REFERENCES

\vskip 0.3in

[C] A. Connes: \it Noncommutative Geometry. \rm Academic Press,
San Diego (1994)

[CDMP]  A. Cardona, C. Ducourtioux, J.P. Magnot, and S. Paycha:
 Weighted traces on the algebra of pseudo-differential operators and
geometry of loop groups. math.OA/0001117

[CZ] G.  Cognola, S. Zerbini: Consistent, covariant and multiplicative
anomalies.  Lett.Math.Phys.\bf 48, \rm 375-383 (1999);  
hep-th/9811039

[D] C. Ducourtioux: Weighted traces of pseudo-differential operators
and associated determinants. Ph.D. thesis, Mathematics Department,
Universit\'e Blaise Pascal, 2001 

[ECZ] E. Elizalde, G. Cognola, and  S. Zerbini: Applications in physics of
the multiplicative anomaly formula involving some basic differential operators. 
Nucl.Phys.\bf B532, \rm 407-428,1998; 
hep-th/9804118 

[EFVZ] E. Elizalde, A. Filippi, L. Vanzo, and  S. Zerbini: Is the
multiplicative anomaly relevant?  
hep-th/9804072 

[F] L. Friedlander: PhD Thesis, Department of Mathematics, MIT (1989)

[F-Sh] L. Faddeev and S. Shatasvili: Algebraic and Hamiltonian methods
in the theory of nonabelian anomalies. Theor. Math. Phys. \bf 60, \rm
770 (1985)

[KV] M. Kontsevich and  S. Vishik: Determinants of elliptic
pseudo-differential operators. hep-th/9404046

[LM] E. Langmann and J. Mickelsson: $(3+1)$-dimensional Schwinger
terms and non-commutative geometry. Phys. Lett. \bf B 338, \rm 241 (1994)

[LM2] E. Langmann and J. Mickelsson:  Elementary derivation of the
          chiral anomaly.
          Lett. Math. Phys. \bf 6, \rm 45 (1996)

[LMR] E. Langmann, J. Mickelsson, and S. Rydh: Anomalies and Schwinger
terms in NCG field theory models. 
J.Math.Phys.\bf 42,\rm 4779, (2001); hep-th/0103006. 

[M] J. Mickelsson: Wodzicki residue and anomalies of current
     algebras. In: \it Integrable Models and Strings. \rm
     Springer LNP 436, p. 123 (1994). Ed. by A. Alekseev,
     A. Hietam\"aki, K. Huitu, and A. Niemi

[M2] J. Mickelsson: Chiral anomalies in even and odd dimensions. Commun. Math. Phys.
     \bf 97, \rm 361 (1985)

[MR] J. Mickelsson and S. Rajeev:  Current algebras in $d+1$
dimensions and determinant bundles over infinite-dimensional
Grassmannians. Commun. Math. Phys. \bf 116, \rm 365 (1988)

[MN] R. Melrose and V. Nistor: Homology of pseudo-differential
operators I. Manifolds with boundary. funct-an/9606005

[O] K. Okikiolu: The multiplicative anomaly for determinants of
elliptic operators. Duke Math. J. \bf 79, \rm 723-750 (1995)

[PR] S. Paycha and S. Rosenberg: Curvature of determinant bundles and
first Chern forms.  math.DG/0009172

[PS] A. Pressley and G. Segal: \it Loop Groups. \rm Oxford University
Press (1986)

[S] B. Simon:  \it Trace Ideals and their Applications. \rm London
Mathematical Society Lecture Notes Series \bf 35, \rm Cambridge
University Press, Cambridge - New York (1979)

\enddocument